\documentclass[]{proarticle}
\usepackage{multicol}
\usepackage{amsmath,amsfonts,amssymb}
 \usepackage[debug,pageanchor=false]{hyperref}
\hypersetup{colorlinks=true,linktocpage,breaklinks,
            urlcolor=blue,
            linkcolor=red,
            citecolor=blue
            }

\def\bea{\begin{eqnarray}}
\def\eea{\end{eqnarray}}
\def\a{\alpha}  \def\g{\gamma} \def\d{\delta} \def\e{\epsilon}
\def\ve{\varepsilon}   
   \def\l{\lambda} \def\m{\mu}
\def\n{\nu} \def\x{\xi} \def\p{\pi}  \def\r{\rho}
 \def\s{\sigma}   \def\f{\varphi}
 \def\c{\chi} \def\y{\psi} \def\w{\omega}

\def\G{\Gamma} 
\def\D{\Delta} 
   
   \def\L{\Lambda} 
 \def\X{\Xi}  
   
\def\F{\Phi}   \def\W{\Omega}

\def\bM{\bar{M}}

\def\fr{\frac}  \def\dt{\partial}

\def\N{\mathcal{N}}

\def\mc{\mathcal}

\def\XX{\mathbb{X}}

\def\RR{\mathbb{R}}
\def\TT{\mathbb{T}}

\def\PP{\mathbb{P}}

\def\beq{\begin{equation}}
\def\eeq{\end{equation}}
\def\bea{\begin{eqnarray}}
\def\eea{\end{eqnarray}}

\def\F{{\mathcal{F}}}

\begin{document}
\title{Exceptional Field Theory for $E_{6(6)}$ supergravity.}
\author{E. Musaev}
\maketitle
\address{National Research University Higher School of Economics, Faculty of
Mathematics
                              \\ 7, st. Vavilova, 117312, Moscow, Russia. }
\eads{emusaev@hse.ru} 
\begin{abstract}
A brief description of the supersymmetric and duality covariant approach to
supergravity is presented. The formalism is based on exceptional geometric
structures and turns the hidden U-duality group into a manifest gauge
symmetry. Tensor hierarchy of gauged supergravity appears naturally here as a
consequence of covariance of the construction. Finally, the full
supersymmetric Lagrangian is explicitly constructed. This work was presented on
the
International Conference ``Quantum Field Theory and Gravity (QFTG'14)'' in
Tomsk.
\end{abstract}


\begin{multicols}{2}
\section{Introduction}

\subsection{Dualities in supergravity}

Since the seminal work of Cremmer and Julia \cite{Cremmer:1978ds} it is well 
known that 11-dimensional supergravity compactified on a torus $\TT^d$
enjoys a hidden
symmetry $E_{d(d)}$. From the point of view of the underlying M-theory these are
the so-called U-duality transformations that unify the perturbative T-duality,
that relates Type IIA and Type IIB theories, and S-duality of Type IIB
string theory.

To get the basic idea of the construction it is the
most instructive to start with $D=11$, $\N=1$ supergravity, whose field content
is very simple. This introduction mainly follows the paper \cite{Cremmer:1979up}
by Cremmer and Julia that contains very clear and detailed review of their
results presented in the letter \cite{Cremmer:1978ds}. The field content of
eleven-dimensional supergravity is very simple: graviton, 11-dimensional
gravitino and the 3-form gauge field. Upon reduction on a d-dimensional torus
$\TT^d$, parametrised by the coordinates $\{x^n\}$, the theory fits into the
maximal supergravity in $D=11-d$ dimensions. Decomposing the 11-dimensional
fields under the split 11=D+d one gets the following field content in 4
dimensions. From the vielbein we get one $D$-dimensional vielbein
$e^{\bar{\alpha}}_\m$, $d$ vector fields $A_\m^m$ and $d(d+1)/2$ scalar fields
$g_{mn}$. The 3-form field reduces into a 3-form $C_{\m\n\r}$, $d$ number of
2-forms $B_{\m\n m}$, $d(d-1)/2$ vectors $A_{\m mn}$ and $q=d(d-1)(d-2)/6$
scalar fields $C_{mnk}$.

Such constructed effective theory has in general $SL(d)\ltimes \RR^q$ global 
(rigid) symmetry group, where the $SL(d)$ part comes from the  diffeomorphisms
of the internal space of the form $\d x^m=\L^m_n x^n$. The abelian group
$\RR^q$, that is the remnant of the gauge symmetry, acts on the axions $C_{mnk}$
as constant shifts
\begin{equation}
\d C_{mnk}=c_{mnk}(=\mbox{const}).
\end{equation}
In addition, in dimensions $D=3,4,5$ one can dualize 1,2 and 3-forms respectively to obtain addition scalars when the $p$-forms enter the Lagrangian only by their derivatives. There are certain subtleties when this procedure is applied to the 11-dimensional supergravity because of the Chern-Simons-like terms $F[C]\wedge F[C] \wedge C$, which will not be described here. Very detailed inspection of the global rigid symmetries that survive this construction is presented in \cite{Cremmer:1997ct}. To be mentioned is that such dualisations are necessary in $D\leq 5$ to obtain the full U-duality group $E_{d(d)}$ in the scalar sector.

Hence, the scalar fields can be nicely packaged into a matrix 
$\mc{V}$ that is an element of the coset
$E_{d(d)}/K(E_{d(d)})$. By choosing a correct parametrisation of the coset the
scalar potential can be written in the following form that is globally invariant
under $E_{d(d)}$ \cite{Cremmer:1997ct}
\begin{equation}
\mc{L}_{scalar}=\fr14eTr[\dt \mc{M}^{-1}\dt\mc{M}],
\end{equation}
where $\mc{M}=\mc{V}^*\mc{V}$ is the metric on the coset space.  The involution
${}^*$ here denotes the usual transposition for $D\geq 6$ and is replaced by
Hermitian conjugate and contraction with a certain symplectic matrix $\W$ for
$D\leq 5$ (this is known as Cartan involution).

It is possible to repeat the same story for the $p$-forms sector,  taking into
account that to have the global symmetry on the level of Lagrangian (not the
EOM), in even dimensions $D=2n$ one has to add extra ''magnetic'' duals to
$n$-forms. This is necessary since on the level of equations of motion the
symmetry is realised on the field strengths rather than the gauge potentials.
Hence, an
$n$-form field strength together with its Hodge dual forms a representation of
the duality group. 

It is important that the hidden symmetries in the described construction 
are global symmetries of a $D$-dimensional effective theory. Following analogy
with General Relativity one may ask what is the geometric origin of the duality
symmetries and to what extent do they present in the initial 11-dimensional
supergravity. The formalism of Exceptional Field Theory that is an attempt to
make sense of these questions and to find a way to answer them is briefly
described in this letter. For calculational details and more involved discussion
the reader may refer to \cite{Musaev:2014ab}.

\subsection{Basic conventions}

In what follows we focus on the $E_6$ exceptional field theory and hence it is
useful to list few basic conventions and definitions that will be used
\cite{deWit:2004nw}. A coset
representative is denoted as usual by 
\begin{equation}
\mc{V}^{ij}_M\in \fr{E_{6(6)}}{USp(8)},
\end{equation}
where the index convention is the following
\begin{equation}
\begin{aligned}
M,N,O,P,\ldots & =1,\ldots, 27, && E_{6(6)} \mbox{ indices}\\
A,B,C,D,\ldots & =1,\ldots, 27, &&\mbox{local } USp(8) \\
i,j,k,\ldots & = 1,\ldots,8,&&\mbox{local } USp(8)\\
\m,\n,\r,\s,\ldots & = 1,\ldots,5,&& GL(5) \mbox{ indices}\\
a,b,c,d,\ldots & = 1,\ldots,5,&&\mbox{local } SO(1,4).
\end{aligned}
\end{equation}
The scalar matrix $\mc{V}_M^{ij}$ and the symplectic $USp(8)$ matrix $\W_{ij}$
satisfy a set of constraint
\begin{equation}
\begin{aligned}
&\mc{V}_M^{ij}\mc{V}^N_{ij}=\d_M^N, && \mc{V}_M^{kl}\mc{V}^m_{ij}=\d^{kl}_{ij}
-\fr18\W_{ij}\W^{kl},\\
&\mc{V}_M^{kl}\W_{kl}=0,&&
\mc{V}_{Mij}=(\mc{V}_M^{ij})^*=\mc{V}_M^{kl}\W_{ki}\W_{lj},\\
&\W_{kl}\W^{lm}=-\d^m_k,
\end{aligned}
\end{equation}
where the star denotes complex conjugation and the Kronecker symbol for pairs of
antisymmetric indices is defined as
$\d^{ij}_{kl}=1/2(\d^i_k\d^j_l-\d^i_l\d^j_k)$. In addition we use the convention
that all (anti)symmetrisations of $n$ indices are performed with a prefactor of
$1/n!$, i.e.
\begin{equation}
A_{[i_1,\ldots i_n]}\equiv \fr{1}{n!}\left(A_{i_1\ldots
i_n}+\mbox{permutations}\right).
\end{equation}
For the spinor sector we use symplectic Majorana spinors $\y^i$ subject to the
reality constraint
\begin{equation}
\begin{aligned}
&C^{-1}\bar{\y}_i^T=\W_{ij}\y^j,&& {\y^i}^TC=\W^{ij}\bar{\y}_j,
\end{aligned}
\end{equation}
where the charge conjugation matrix $C$ is defined by the following relations
\begin{equation}
\begin{aligned}
C\g_aC^{-1}=\g_a^T, && C^T=-C, && C^\dagger = C^{-1} \\
 \g_{abcde}=\mathbf{1}\ve_{abcde}.
\end{aligned}
\end{equation}
This implies the following relation for fermionic bilinears with spinor fields
$\y^i$ and $\f^i$ 
\begin{equation}
\bar{\y}_i\G \f^j=-\W_{ik}\W^{jl}\bar{\y}_l(C^{-1}\G^TC)\y^k
\end{equation}
for any expression of gamma matrices $\G$.

\subsection{Extended geometry}

Following the construction of Cremmer and Julia the
hidden exceptional symmetries of lower dimensional maximal supergravities
most straightforwardly can be reproduced in toroidal reductions of
11-dimensional supergravity. The formalism of extended geometry provides more
geometric background to the exceptional groups in terms of extended geometric
structures on an extended space (for review see
\cite{Aldazabal:2013sca,Berman:2013eva,Hohm:2013bwa}).

The extended space is constructed by adding extra directions to the would-be
internal manifold that correspond to winding modes of M-branes
\cite{Hull:2007zu,Hohm:2010pp}
\begin{equation}
 \XX^M=\{x^m,y_{mn},z_{mnklp},\ldots\}.
\end{equation}
Infinitesimal coordinate transformations on this space consistent with the
exceptional groups are defined as a generalisation of the well-known Hitchin's
construction. Hence, one defines generalised tensors that live on the extended
space as objects with the following transformation rule
\cite{Coimbra:2011ky,Berman:2012vc}
\begin{equation}
\label{gen_lie}
\begin{aligned}
(\mc{L}_\L T)^M=&\L^N\dt_N T^M-6 \PP^M{}_L{}^N{}_K\dt_N \L^K T^L \\
& + \l_T (\dt_K \L^K)T^M\equiv[\L,T]_D^M.
\end{aligned}
\end{equation}
The first and the last terms play the roles of translation and a weight term
respectively. The second term
reflects the exceptional group symmetry and involves the projection of the
matrix $\dt_N \L^K$ on the U-duality algebra,
since in general it does not belong to the structure group $E_{6(6)}$
\cite{Hohm:2013vpa}. This is
very similar to General Relativity where, however, the group is $GL(n)$ and any
non-degenerate matrix belongs to its algebra. Hence, in the case of the  $GL$
geometry the projector will be just trivial.

In addition one introduces a differential constraint on all fields in  the
theory that restricts dependence on the extended coordinates $\XX^M$
\begin{equation}
d^{PMN}\dt_M\otimes \dt_N=0.
\end{equation}
This extra condition in particular implies existence of a trivial transformation
given by $\L_0^M=d^{MNK}\dt_N\X_{K}$ which itself transforms
as a generalised vector. The Jacobi identity and closure of the algebra  hold up
to a trivial transformation as well. The latter leads to
the notion of E-bracket that is an
antisymmetrisation of the Dorfman bracket
\begin{equation}
\begin{aligned}
{[}\mc{L}_{\L_1},\mc{L}_{\L_2}{]}&=\mc{L}_{[\L_1,\L_2]_E},\\
[\L_1,\L_2]_E&\equiv[\L_{[1},\L_{2]}]_D.
\end{aligned}
\end{equation}
It is important to mention that in contrast to the E-bracket, the Dorfman
bracket $[,]_D$ is not antisymmetric nor symmetric. This will play a crucial
role in
construction of tensor hierarchy starting from the covariant derivative to be
defined in the next section.

\section{$E_{6(6)}$ covariant exceptional field theory}

\subsection{Covariant derivative for D-bracket and tensor hierarchy}

In the formalism of Extended Geometry generalised tensors and the corresponding
transformations are considered to be independent of space-time coordinates
$x^\m$, that decouples the would be scalar sector. 

In order to naturally incorporate the tensor and fermionic sector into the
formalism the fields and all the gauge parameters are now allowed to depend on
the external space-time coordinates. In the spirit of the ordinary
Yang-Mills construction this implies that one has to introduce a
long space-time derivative, that is covariant with respect to D-bracket
\cite{Hohm:2013vpa}
\begin{equation}
\begin{aligned}
\d_\L D_\m T^M&=\mc{L}_\L D_\m T^M,\\
{D}_\m&=\dt_\m-\mc{L}_{A_\m^M}=\dt_\m-[A_\m,\,]_D,\\
\d_\L A_\m^M&=\dt_\m \L^M-[A_\m,\L]_D^M=\mc{D}_\m\L^M,
\end{aligned}
\end{equation}
where the gauge field $A_\m^M$ is identified with the vector field of the
corresponding maximal supergravity (with all necessary dualisations).

Since the E-bracket does not satisfy the Jacobi identity,
one has to deform the usual field strength by a trivial transformation 
\begin{equation}
\begin{aligned}
{[}\mc{D}_\m,\mc{D}_\n{]}&=-\mc{L}_{\F_{\m\n}},\\
\F_{\m\n}^M&=2\dt_{[\m}A_{\n]}^M-[A_{\m},A_{\n}]_E^M+10d^{MNK}\dt_NB_{K\m\n}.
\end{aligned}
\end{equation}
In the spirit of tensor hierarchy, gauge transformations of the 1- and 2-forms
naturally have the following form
\begin{equation}
\label{trans_AB}
\begin{aligned}
\d A_\m^M&=\mc{D}_\m\L^M-10d^{MNK}\dt_N\X_{K\m},\\
\D
B_{M\m\n}&=2\mc{D}_{[\m}\X_{M\n]}+d_{MNK}\L^N\mc{F}_{\m\n}{}^K.
\end{aligned}
\end{equation}

This construction nicely utilises the $p$-forms of the $D=5$ maximal
supergravity and naturally leads to tensor hierarchy as a consequence of
generalised covariance.

\subsection{Geometry and connections}

The structure of EFT explicitly distinguishes between the two  sets of
coordinates: space-time $\{x^\m\}$ and the extended space $\XX^M$. Respectively,
one has two local groups $SO(1,4)$ and $USp(8)$. Hence, there are four
types of connections listed in the following table

\begin{center}
\begin{tabular}{c|ccc}
& $\mc{D}_\m$ & $\nabla_M$ \\
\hline \\[-0.3cm]
$SO(1,4)$ & $\w_\m{}^{ab}$  & $\w_M{}^{ab}$ \\[0.15cm]
$USp(8)$ & $Q_\m{}^i{}_j$ & $Q_M{}^i{}_j$
\end{tabular}
\end{center}

The $SO(1,4)$ connection $\w_\m{}^{ab}$ is defined by the usual vanishing
torsion
condition $\mc{D}_{[\m} e_{\n]}^a =0$. The $USp(8)$ connection $Q_\m$ is
defined according to the group properties of the matrix $\mc{V}_M^{ij}$ as
usual (see \cite{deWit:2004nw})
\begin{equation}
\label{current}
\begin{aligned}
& \mc{V}_{kl}{}^M
D_\m\mc{V}_M{}^{ij}=2\d_k{}^{[i}Q_\m{}_l{}^{j]}+P_\m{}^{ijmn}\W_{mk}\W_{nl},\\
& Q_\m \in usp(8), \quad P_\m \in \mathfrak{e}_{6}\ominus usp(8).
\end{aligned}
\end{equation}
Explicit form of the $SO(1,4)$ connection $\w_M{}^{ab}$ can
be found following the same story but for the space-time vielbein $e_\m^a$, i.e.
\begin{equation}
e^{a\m} \dt_M e_\m^b = \w_M{}^{ab}+\p_M{}^{ab},
\end{equation}
where $\p_M{}^{ab}=\p_M{}^{ba}$. Finally, the internal $USp(8)$ connection
$Q_M{}^i{}_j$ is derived from an analogue of the
vanishing torsion condition for the extended space vielbein $\mc{V}_M^{ij}$. The
generalised torsion is given by 
\begin{equation}
\mc{T}_{NK}{}^M=\G_{NK}{}^M-6\mathbb{P}^M{}_K{}^P{}_L\G_{PN}{}^L+\fr32\mathbb{P}
^M{}_K{}^Q{}_N\G_{PQ}{}^P,
\end{equation}
that follows from the usual definition
$\mc{T}(V,W)^M=\mc{L}^\nabla_VW^M-\mc{L}_VW^M$,
where $\mc{L}^\nabla$ is the covariant generalised Lie derivative. 
Hence one may write for the
vanishing torsion condition
\begin{multline}
\mc{V}_{\bM}{}^M\mc{D}_N\mc{V}_K{}^{\bM}-6\mathbb{P}^M{}_K{}^P{}_L\mc{V}_{\bM}{}
^L\mc{D}_P\mc{V}_N{}^{\bM}\\ +\fr32\mathbb{P}^M{}_K{}^Q{}_N\G_{PQ}{}^P=0.
\end{multline}
This equation has the form of the familiar expression $\nabla_{[\m}
e_{\n]}^a=0$ however
deformed in accordance to the algebraic structure of the duality group $E_{6}$.

\section{Supersymmetry transformations}
\label{susy}

Supersymmetry transformations of the fields of
$E_{6(6)}$ covariant supergravity are taken to be of the following form
\end{multicols}
\begin{equation}
\label{transf}
\begin{aligned}
\d_\e e^a_\m&=\fr12\bar{\e}_i\g^a\y_\m^i, \quad \d_\e \y_\m^i=\mc{D}_\m
\e^i-i\sqrt{2}\mc{V}^{Mij}\left(\tilde{\nabla}_M(\g_\m\e^k)-\fr13\g_\m\tilde{
\nabla }
_M\e^k\right)\W_{jk},\\
\d_\e\c^{ijk}&=\fr i2 P_\m^{ijkl}\g^\m\W_{lm}\e^m-
\fr{3}{\sqrt{2}}\left(\mc{V}^{M[ij}\W^{k]m}-\fr13\mc{V}^{M
m[i}\W^{jk]}\right)\W_{mr}\tilde{\nabla}_M\e^r,\\
\d_\e\mc{V}_M^{ij}&=i\mc{V}_M{}^{kl}\Big[4\W_{p[k}\bar{\c}_{lmn]}\e^p+3\W_{[kl}
\bar{\c}_{mn]p}\e^p\Big]\W^{mi}\W^{nj},\\
\d_\e
A_\m{}^M&=\sqrt{2}\Big[i\W^{ik}\bar{\e}_k\y_\m{}^j+\bar{\e}_k\g_{\m}\c^{ijk}\Big
] \mc {
V}_{ij}{}^M,\\
\d_\e
B_{\m\n}{}_M&=\fr{1}{\sqrt{5}}\mc{V}_{M}{}^{ij}\Big[2\bar{\y}_{i[\m}\g_{\n]}
\e^k\W_{jk}-i\bar{\c}_{ijk}\g_{\m\n}\e^k\Big]+d_{MNP}A_{[\m}{}^N\d_\e,
 A_{\n]}^P
\end{aligned}
\end{equation}
\begin{multicols}{2}
where we define the full covariant derivative as
\begin{equation}
\begin{aligned}
\tilde{\nabla}_M\e^i&=\nabla_M\e^i-\fr{1}{8}\mc{F}_M{}^{\r\s}\g_{\r\s}\e^i.
\end{aligned}
\end{equation}

Closure of supersymmetry transformations  on the
fields of EFT has the following structure
\begin{equation}
\begin{aligned}
[\d(\e_1),\d(\e_2)]&=\x^\m\mc{D}_\m+\d_{so(1,4)}(\W^{ab})+\d_{usp(8)}(\L^{ij}
)\\
&+\d_{gauge}(\L^M)+\d_{gauge}(\X_{M\m})\\
&+\d_{gauge}(\X^\a{}_{\m\n})+\d_{susy}
(\e_3)+\d(\mc{O}_{M\m\n}),
\end{aligned}
\end{equation}
that is the
same as for the five-dimensional theory.
Parameters of the transformation on the RHS are given by the following
expressions made of the spinors $\e_{1,2}$ and the scalar matrix
$\mc{V}_M{}^{ij}$
\begin{equation}
\begin{aligned}
&\x^\m=\fr12\bar{\e}_{2i}\g^\m\e_1^i,\quad 
\L^M=-\fr{i}{\sqrt{2}}\left(\bar{\e}_{2i}\e^k_1\mc{V}^{Mij}\W_{jk}\right),\\
 &\W^{ab}=-\fr{i\sqrt{2}}{3}\left(\bar{\e}_{1i}\g^{ab}\tilde{\nabla}_M\e_2^k-
\tilde{\nabla}_M\bar{\e}_{1i}\g^{ab}\e_2^k\right)\mc{V}^{Mij}\W_{jk}\\
&-\L^M\w_M{}^{ab}+\fr{1}{2}\L^M\mc{F}_M{}^{ab},\\
&\X_{M\m}=-\fr{1}{\sqrt{5}}\mc{V}_M{}^{kl}\W_{lm}\big(\bar{\e}_{2k}
\g_\m\e_1^m\big)\\
&\X^\a{}_{\m\n}=-\fr{3i}{\sqrt{10}}(t^\a)^M{}_N\mc{V}_{Msi}\mc{V}^{Nki}
\big(\bar
{\e}_{2k}\g_{\m\n}\e_1^s\big),\\
&\mc{O}_{M\m\n}=\fr{i}{\sqrt{10}}\Big(\bar{\e}_{2k}\g_{\m\n}
\dt_M\e_1^k-\dt_M\bar{\e}_{2k}\g_{\m\n}\e_1^k\\
&-(\bar{\e}_{2k}\e_1^s)
e^a_{[\m}\dt_Me_{\n]a}+\fr23\mc{V}_{Nsi}\dt_M\mc{V}^{Nki}\big(\bar{\e}_{2k}\g_
{\m\n}\e_1^s\big)\Big).
\end{aligned}
\end{equation}
Here $\x^\m$ and $\L^M$ are the diffeomorphism parameters, $\W^{ab}$
parametrizes the
Lorentz rotations, $\X_{M\m}$ and $\X_M^\a$ are the gauge
transformation parameters of the 2-form $B_{\m\n K}$ and the extra 3-form
$C^{\a}{}_{\m\n\r}$. Finally, as a consequence of the section condition the
tensor $\mc{O}_{M\m\n}$ is constrained by
\begin{equation}
d^{MNK}\dt_N\mc{O}_{K\m\n}=0.
\end{equation}
The operator 
$\d(\mc{O}_{M\m\n})$ leaves invariant the field
$\mc{F}_{\m\n}{}^M$ that is the only way of how the
2-form field enters the Lagrangian. The same is true for the gauge
transformation generated by $\X^\a{}_{\m\n}$. Hence, the superalgebra is
closed up to the section condition.

\section{Invariant Lagrangian}

Given the definitions of the covariant derivatives that respect the $E_{6(6)}$
structure of the extended space the full supersymmetric Lagrangian for the
covariant Exceptional Field Theory takes the following form
\begin{equation}
\begin{aligned}
\label{L}
&e^{-1}\mc{L}=R-\fr{1}{4}\,\mc{M}_{MN}\,\mc{F}_{\m\n}{}^{M}\mc{F}^{\m\n
N}-\fr{1}{6}\,|\mc{P}_\m{}^{ijkl}|^2\\
&-\bar{\y}_{\m
i}\g^{\m\n\r}\mc{D}_\n\y_\r^i-2\sqrt{2}i\mc{V}^M{}_{ij}\W^{ik}\bar{\y}_{\m
k}\g^{[\m}\tilde{\nabla}'_M(\g^{\n]}\y_\n{}^j)\\
&{}-\fr{4}{3}\,\bar{\c}_{ijk}\g^\m\mc{D}_\m\c^{ijk}-4\sqrt{2}i\mc{V}
^M
{}_{mn}\W^{np}\bar{\c}_{pkl}\tilde{\nabla}'_M\c^{mkl}\\
&{}+\fr{4i}{3}\mc{P}_\m{}^{ijkl}\bar{\c}_{ijk}\g^\n\g^\m
\y_\n{}^m\W_{lm}\\
&+4\sqrt{2}\mc{V}^{Mij}\bar{\c}_{ijk}
\g^\m\tilde {\nabla}_M\y_\m{}^k+\mc{L}_{\rm top}-V(\mc{M},g),
\end{aligned}
\end{equation}
where $\tilde{\nabla}'[\mc{F}]=\tilde{\nabla}[-\mc{F}]$ encodes switch of the sign of
the gauge field flux, $\mc{L}_{\rm top}$ is the topological term that includes the covariant
version of the Chern-Simons Lagrangian and $V$ is the scalar potential of
\cite{Hohm:2013vpa,Musaev:2013rq}.
Due to the lack of space we do not check explicitly supersymmetry
invariance of the above Lagrangian. For more detailed consideration the reader
is referred to the paper \cite{Musaev:2014ab}. However, the
$E_{6(6)}$ invariance is
 manifest since all the objects in the Lagrangian are covariant.

\section{Outlook}

In this note the U-duality covariant approach to supergravity is briefly
described. The essential feature of the Exceptional Field Theory approach is
the notion of extended space and the structure of extended geometry defined on
it. We describe the construction of $E_{6(6)}$ covariant derivatives in
both the space-time and the extended space directions. The corresponding
vanishing torsion and algebraic conditions give necessary expressions
for the $SO(1,4)$ and $USp(8)$ connections.

The final result is the supersymmetric manifestly $E_{6(6)}$-covariant
Lagrangian that includes all the fields of the maximal $D=5$ supergravity. The
11-dimensional diffeomorphism symmetry is not manifest in this construction,
however upon solution of the section constraint one is able to restore the
full 11- or 10-dimensional Lagrangian.

As it was shown in \cite{Hohm:2013vpa}, decomposition of the $27$ extended space
coordinates under the $GL(6)$ subgroup of $E_{6(6)}$ and leaving only the
coordinates in the {\bf 6}, provides a consistent solution of
the section constraint. This corresponds to the Kaluza-Klein decomposition
of the full 11-dimensional supergravity. 

An alternative solution is given by decomposition of the {\bf 27} under
the $GL(5)\times SL(2)$ subgroup. This leads to Type IIB supergravity with
manifest $SL(2)$ duality symmetry.

Relation between the described
formalism and the embedding tensor approach to gauged supergravities is
given by generalised Scherk-Schwarz reductions
\cite{Grana:2012rr}. As it was shown in
\cite{Berman:2012uy,Musaev:2013rq} the reduction naturally provides all the
gaugings in terms of
generalised twist matrices and their derivatives with respect to the full set
of extended coordinates. 

\section*{Acknowledgement}
The author expresses his gratitude to theoretical dpt of CERN for warm
hospitality during completion of this letter. In addition I would like to
thank ENS de Lyon and personally Henning Samtleben for generous
financial support and productive collaboration. Finally, I thank
Tomsk State
Pedagogical University and personally Vladimir Epp and Joseph Buchbinder for
creating a wonderful atmosphere during the QFTG'2014 conference. 
\end{multicols}

\end{document}